\documentclass[]{article}
\usepackage[style=chem-acs,articletitle=true]{biblatex}
\usepackage[parfill]{parskip}

\usepackage{geometry}
\usepackage{float}
\usepackage{siunitx}
\usepackage{etoc}
 \usepackage[font=small,font=sf, labelfont=bf]{caption}

\newcommand{\beginsupplement}{%
        \setcounter{section}{0}
        \renewcommand{\thesubsection}{\arabic{subsection}}%
        \setcounter{table}{0}
        \renewcommand{\thetable}{S.\arabic{table}}%
        \setcounter{figure}{0}
        \renewcommand{\thefigure}{S.\arabic{figure}}%
        \setcounter{equation}{0}
        \renewcommand{\theequation}{S.\arabic{equation}}%
     }

\usepackage[utf8]{inputenc}
\usepackage{mathtools}
\usepackage{amsmath}
\usepackage{braket}
\usepackage{textcomp}
\usepackage{xcolor}
\usepackage{hyperref}

\title{\Large{\textbf{Directional excitation of a high-density magnon gas using coherently driven spin waves}}}
\author{Brecht G. Simon$^{1,\dagger}$, Samer Kurdi$^{1,\dagger}$, Helena La$^1$, Iacopo Bertelli$^{1,2}$, Joris J. Carmiggelt$^1$,\\ Maximilian Ruf$^3$, Nick de Jong$^{3,4}$, Hans van den Berg$^{3,4}$, Allard Katan$^1$, Toeno van der Sar$^{1,*}$}
\date{

\begin{flushleft}
\textbf{Affiliations}\\
\small{
$^1$Department of Quantum Nanoscience, Kavli Institute of Nanoscience, Delft University of Technology, 2628 CJ, Delft, The Netherlands\\
$^2$Huygens-Kamerlingh Onnes Laboratorium, Leiden University, 2300 RA, Leiden, The Netherlands\\
$^3$QuTech, Delft University of Technology, 2628 CJ, Delft, The Netherlands\\
$^4$Netherlands Organisation for Applied Scientific Research (TNO), 2628 CK Delft, The Netherlands\\
\hfill\break
$^\dagger$ These authors contributed equally to this work.\\
$^*$ Corresponding author. Email: t.vandersar@tudelft.nl }
\end{flushleft}}
\begin{document}
\maketitle
\begin{abstract}
    Controlling magnon densities in magnetic materials enables driving spin transport in magnonic devices. We demonstrate the creation of large, out-of-equilibrium magnon densities in a thin-film magnetic insulator via microwave excitation of coherent spin waves and subsequent multi-magnon scattering. We image both the coherent spin waves and the resulting incoherent magnon gas using scanning-probe magnetometry based on electron spins in diamond. We find that the gas extends unidirectionally over hundreds of micrometers from the excitation stripline. Surprisingly, the gas density far exceeds that expected for a boson system following a Bose-Einstein distribution with a maximum value of the chemical potential. We characterize the momentum distribution of the gas by measuring the nanoscale spatial decay of the magnetic stray fields. Our results show that driving coherent spin waves leads to a strong out-of-equilibrium occupation of the spin-wave band, opening new possibilities for controlling spin transport and magnetic dynamics in target directions.
\end{abstract}
\begin{refsection}
\addcontentsline{toc}{section}{Main Text}
Spin waves are collective, wave-like precessions of spins in magnetically ordered materials. Magnons are the bosonic excitations of the spin-wave modes. The ability to control the number of magnons occupying the spin-wave energy band is important for driving spin transport in spin-wave devices such as magnon transistors\cite{cornelissen2015long, cornelissen2018spin, wimmer2019spin, chumak2014magnon}. In addition, the generation of large magnon densities can trigger phenomena such as magnetic phase transitions\cite{nan2020electric}, magnon condensation\cite{schneider2020bose, demokritov2006bose, demokritov2008quantum}, and domain-wall motion\cite{kim2014propulsion, shen2020driving,chang2018ferromagnetic, wang2015magnon}. As such, several methods to control magnon densities have been developed, with key methods including spin pumping based on the spin-Hall effect\cite{cornelissen2015long, cornelissen2018spin, kajiwara2010transmission, schneider2021control}, in which magnons are created by sending an electric current through heavy-metal electrodes, and microwave driving of ferromagnetic resonance (FMR)\cite{wettling1983light,bauer2015nonlinear, mccullian2020broadband, du2017control} via metallic electrodes deposited onto the magnetic films.

Here, we demonstrate how the excitation of coherent, traveling spin-wave modes in a thin-film magnetic insulator can be used to generate a high-density, out-of-equilibrium magnon gas unidirectionally with respect to an excitation stripline. We characterize this process using scanning-probe magnetometry based on spins in diamond, a technique which enables probing magnons in thin-film magnets at microwave frequencies by detecting their magnetic stray fields\cite{mccullian2020broadband,van2015nanometre,casola2018probing, bertelli2020magnetic}. We find that the magnon gas has an unexpectedly high density that far exceeds the density expected for a magnon gas following a Bose-Einstein distribution with the maximum possible value of the chemical potential\cite{pitaevskii2016bose}, opening new opportunities for creating and manipulating magnon condensates\cite{schneider2020bose, demokritov2006bose, demokritov2008quantum, demidov2007thermalization}. We further characterize the gas by probing its momentum distribution through distance-dependent measurements of the stray-field magnetic noise it creates. The observed nanoscale spatial decay lengths reveal the presence of large-wavenumber magnons in the gas and underscores the need for nanometer proximity enabled by our scanning-probe magnetometer.

\begin{figure}[H]
    \centering
    \includegraphics[width=\textwidth]{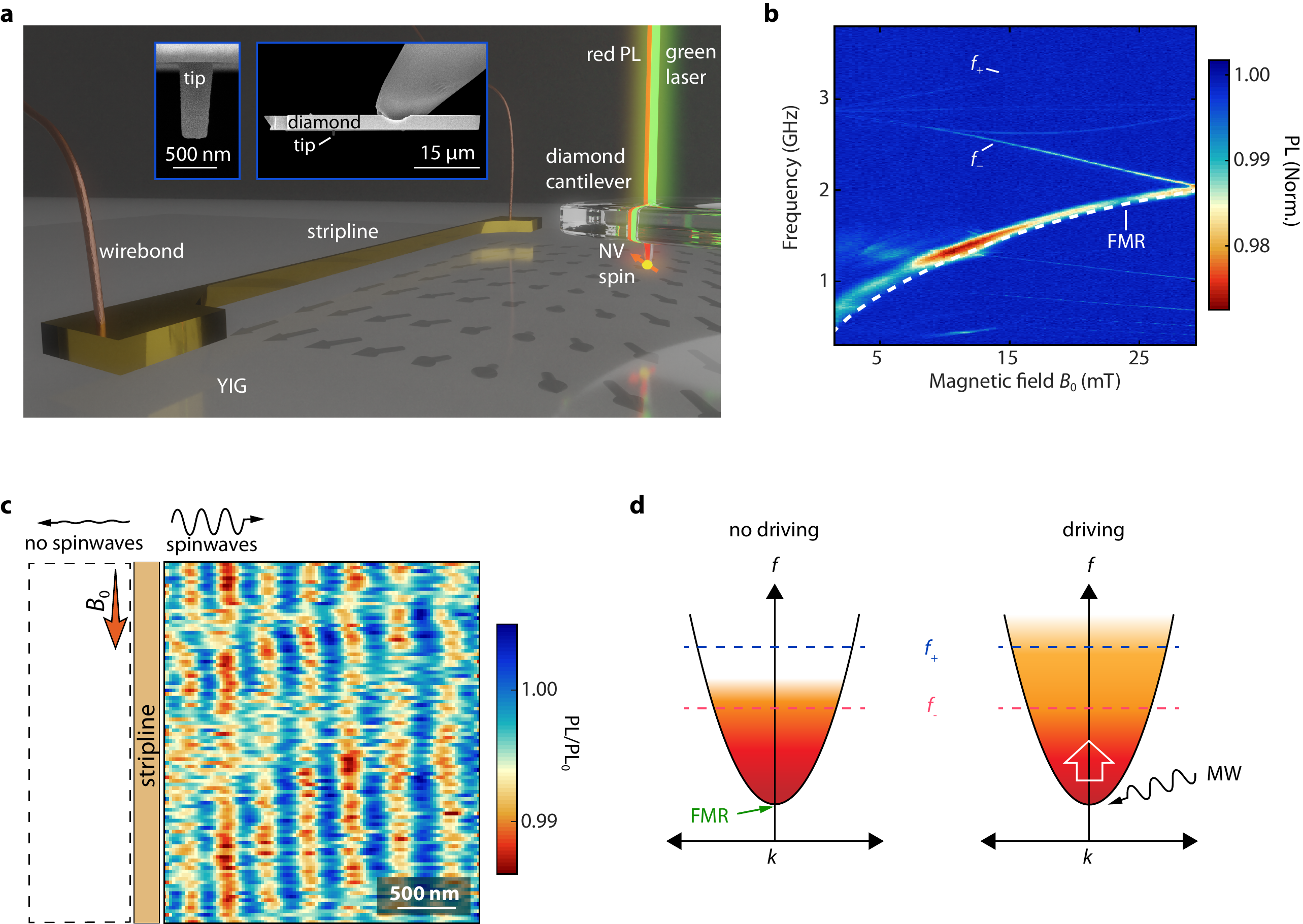}
    \caption{\textbf{Detecting coherent spin waves and incoherent magnon densities using NV spins in diamond.}(a) A diamond cantilever, with NVs implanted $\sim 20$ nm below the tip surface, is mounted in an atomic force microscope (AFM) setup and used for probing the stray-field of spin waves that are excited by a gold stripline. The NV spins are initialized using a green laser and read out via spin-dependent photoluminescence (PL). Insets: scanning electron micrographs of diamond cantilever and tip. (b) Normalized NV photoluminescence vs external field $B_0$ and microwave drive frequency. The ESR frequencies ($f_\pm$), of the NV family that is most aligned with $B_0$ are labelled. The strong PL response close to the ferromagnetic resonance (FMR) is a result of the process depicted in (d). The FMR is calculated as $f_{\mathrm{FMR}}=\gamma\sqrt{B_{\mathrm{IP}}\left(B_{\mathrm{IP}} + \mu_0M_\mathrm{s}\right)}$, where $\gamma = 28 $ GHz/T, $\mu_0 = 4\pi \cdot 10^{-7}$ H/m, $M_\mathrm{s} = 1.42 \cdot 10^5$ A/m and $B_{\mathrm{IP}}$ is the in-plane magnetic field component (Supporting Information Note \ref{Sup:Bfield}). (c) Spatial map of the normalized NV ESR PL showing a coherent spin wave excited unidirectionally (to the right) by applying a microwave current at $f_-$ through the stripline. On the left of the stripline there is no detectable spin-wave signal. The film is magnetized along the stripline direction by a magnetic field $B_0$ (orange arrow), which is set to a low value ($B_0 \sim 0$ mT) in this measurement. At each pixel, the measured PL under microwave driving is normalized to that without microwave driving ($\mathrm{PL}_0$). The image is low-pass filtered to reduce pixel-by-pixel noise. (d) Sketch of the spin-wave dispersion (black line) and its occupation by magnons (color gradient). Without microwave driving (left), only thermally excited magnons are present. Microwave (MW) driving near the FMR frequency (oscillating arrow in right panel) increases the magnon density, which can be detected via the increased stray-field noise at the NV ESR frequencies ($f_{\pm}$).}
    \label{fig:figure1}
\end{figure}

Our scanning-probe magnetometer is based on nitrogen-vacancy (NV) ensembles embedded in the tip of a diamond probe (Fig. \ref{fig:figure1}a)\cite{maletinsky2012robust}(Supporting Information Note \ref{Sup:YIG}-\ref{Sup:TipFab}). The electron spins associated with the NV centers act as magnetic-field sensors that we read out via their spin dependent photoluminescence\cite{casola2018probing}. We use the NV sensors to locally characterize the magnetic stray fields generated by spin waves in a 235-nm thick film of yttrium iron garnet (YIG)\cite{wang2020electrical}, a magnetic insulator with record-long spin-wave lifetimes\cite{serga2010yig}. We employ two measurement modalities to shed light on the interaction between coherently driven spin waves and the resulting out-of-equilibrium magnon gas at higher frequencies: in the first, we measure the coherent NV-spin rotation rate (Rabi frequency) to image the coherent spin waves excited by the stripline. In the second, we drive coherent spin waves with frequencies near the bottom of the spin-wave band while we measure the NV spin relaxation rates at frequencies hundreds of MHz above the drive frequency to characterize the local density of the magnon gas. 

We reveal the directionality of the coherent spin waves launched by the stripline by spatially mapping the contrast of the $f_-$ ESR transition (Fig. \ref{fig:figure1}c). At $B_0\sim0$, this transition is resonant with spin waves of wavelength $\sim$\SI{500}{\nano\meter}, as expected from the known spin-wave dispersion (Supporting Information Note \ref{Sup:RateCalc}). On the right-hand side of the stripline, we observe a spatial standing-wave pattern in the ESR contrast that results from the interference between the direct field of the stripline and the stray fields of the spin waves launched by the stripline\cite{bertelli2020magnetic}. In contrast, we do not observe a spin-wave signal to the left of the stripline. This directionality is characteristic of coherent spin waves traveling perpendicularly to the magnetization and results from the handedness of the stripline field and the precessional motion of the spins\cite{yu2019chiral}.

\begin{figure}[H]
    \centering
    \includegraphics[width=\textwidth]{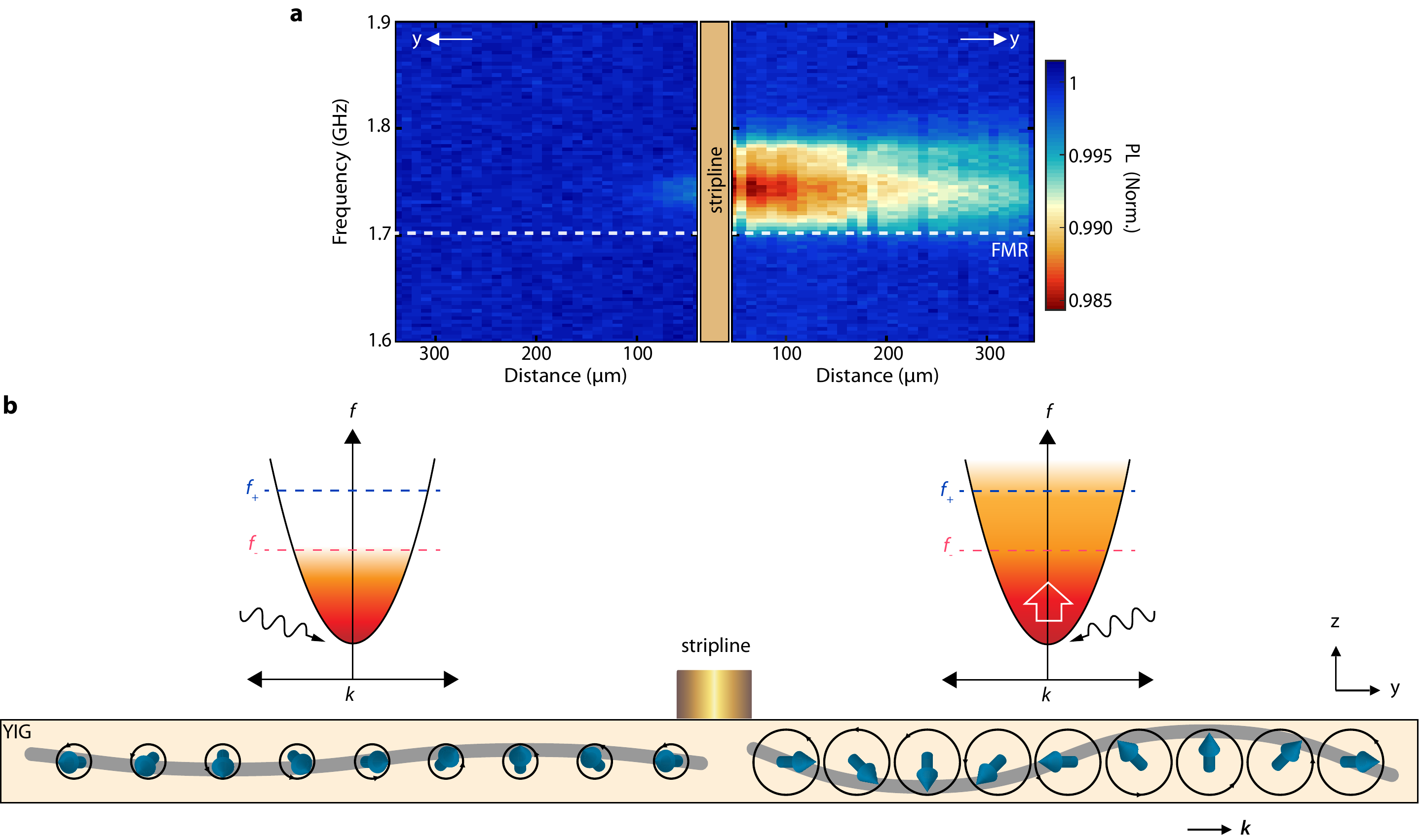}
    \caption{\textbf{Unidirectional excitation of an out-of-equilibrium magnon gas by microwave driving near the ferromagnetic resonance.} (a) One-dimensional spatial maps of the NV photoluminescence (PL) as a function of the frequency of a microwave drive current in the stripline. The microwave current excites directional coherent spin waves (traveling to the right in the image) at near-FMR frequencies. The decrease in NV photoluminescence is a result of incoherent magnons generated at the NV ESR frequencies via multi-magnon scattering (see schematics in (b)). An external magnetic field $B_0=20$ mT magnetizes the film along the stripline direction. The NV ESR frequencies at this field can be seen in Fig. \ref{fig:figure1}b. (b) Schematic of the directional excitation of a coherent spin wave by the microwave stripline and the accompanying out-of-equilibrium occupation of the spin-wave band that arises through multi-magnon scattering. The coherent spin waves are depicted as a collective wave-like precession of the spins in the film (blue arrows). The increased excitation efficiency for right-propagating spin waves results in larger spin-wave amplitudes on the right-hand side (grey lines). The spin-wave band is schematically depicted by a parabolic dispersion. The microwave field that drives the coherent spin waves is indicated by the oscillating arrow. The out-of-equilibrium occupation of the spin-wave band is indicated by the color gradient (as in Fig. \ref{fig:figure1}d.) The NV detection frequencies $f_\pm$ are indicated by dashed lines.}
    \label{fig:figure2}
\end{figure}

In addition to the narrow lines of reduced photoluminescence indicating the NV ESR frequencies (Fig. \ref{fig:figure1}b), we observe a broad band of photoluminescence reduction close to the expected ferromagnetic resonance (FMR) frequency of our YIG film that is detuned from the NV ESR transitions. A similar off-resonant NV response was observed previously\cite{mccullian2020broadband, du2017control, van2015nanometre,wang2020electrical, wolfe2014off, wolfe2016spatially, lee2020nanoscale}, and has been attributed to the driving of a uniform FMR mode and subsequent multi-magnon scattering. The scattering processes lead to an increased magnon density at the NV ESR frequencies, causing NV spin relaxation\cite{mccullian2020broadband, du2017control} (Fig. \ref{fig:figure1}d). However, in contrast with the uniform nature of the FMR mode, we observe that the signal strength depends strongly on the detection location with respect to the stripline (Fig. \ref{fig:figure2}a): on the right-hand side, we observe a much stronger response than on the left, up to distances \SI{300}{\micro\meter} away from the stripline. This asymmetry shows that directional spin waves excited by the stripline, such as those in Fig. \ref{fig:figure1}c, underlie the increased magnon densities at the NV ESR frequencies (Fig. \ref{fig:figure2}b).

\begin{figure}[H]
    \centering
    \includegraphics[width = \textwidth]{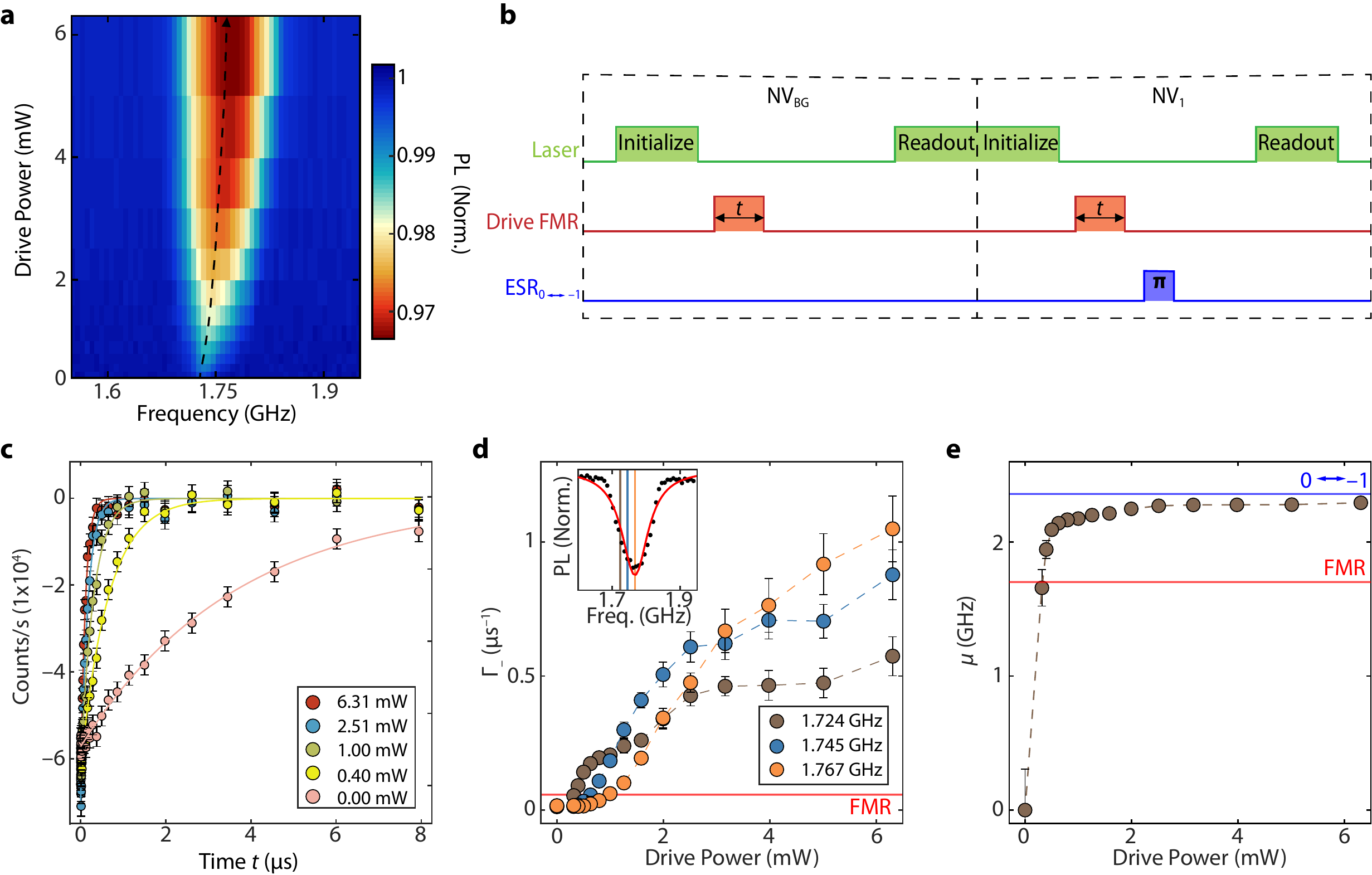}
    \caption{\textbf{Characterizing the density of the magnon gas under near-FMR driving using NV relaxometry. }(a) Normalized NV PL versus microwave drive frequency and power. The maximum contrast shifts with drive power (dashed black line). (b) Measurement sequence to characterize the magnon density at the $f_-$ NV ESR frequency: a \SI{2}{\micro\second} laser pulse prepares the NV spins in $\ket{0}$. A variable-duration microwave pulse (red) causes an out-of-equilibrium magnon density that induces NV spin relaxation. Two sequences are performed, with and without a microwave $\pi$-pulse. The $\pi$-pulse at the $f_-$  ESR frequency switches the $\ket{0}$ and $\ket{-1}$ populations of the target NV spins. The final spin state is characterized by measuring the PL during the first 600 ns of a green laser pulse. The PL difference between the two sequences enables the extraction of the $\ket{0} \leftrightarrow \ket{-1}$ relaxation rate of the target NV spins assuming negligible $\ket{0} \leftrightarrow \ket{+1}$⟩ relaxation (Supporting Information Note \ref{Sup:RateMeas}). (c) Relaxation rate measurement of the PL difference between the two sequences versus duration of the spin-wave drive pulse (see (b)), for different drive powers. $B_0=20$ mT. Filled circles: data. Solid lines: fits to $Ae^{\left(-2\Gamma_- t\right)}$, where $A$ is an offset and $\Gamma_-$  the $\ket{0} \leftrightarrow \ket{1}$  relaxation rate. The error bars represent uncertainties stemming from photon shot noise. (d) $\Gamma_-$  versus drive power for three near-FMR drive frequencies, extracted from measurements as in (b). The red line labelled ‘FMR’ indicates the maximum rate for a system described by a chemical potential that saturates at the bottom of the spin-wave band. Error bars determined from fits. Inset: Normalized NV PL versus microwave drive frequency at 6.3 mW microwave power (see (c)). The vertical lines indicate the drive frequencies used for the measurements in the main panel. (e) Effective chemical potential as a function of drive power. Blue line: ESR transition frequency (2.359 GHz). Red line: Calculated FMR frequency (1.702 GHz). }
    \label{fig:figure3}
\end{figure}

Next, we study the density of the magnon gas created via the driving of directional spin waves. Magnons can re-distribute over the spin-wave band through magnon-magnon interactions and lead to an equilibrated occupation described by a Rayleigh-Jeans distribution\cite{demokritov2006bose} with chemical potential $\mu$\cite{demidov2007thermalization}: $n(f, \mu) = \frac{k_\mathrm{B} T}{hf-\mu}$ (which is the high-temperature limit of the Bose-Einstein distribution, appropriate for our room-temperature measurements), here $k_B$ is Boltzmann’s constant, $T$ is the temperature, $h$ is Planck’s constant and $f$ is the probe frequency. To study whether the magnon gas (Fig. \ref{fig:figure2}a) is described by this distribution we monitor the magnon density at the $f_-$ ESR frequency while driving directional spin waves. To determine which drive frequency yields the strongest NV response, we first characterize the NV photoluminescence while sweeping the frequency and power of the microwave drive field (Fig. \ref{fig:figure3}a). Then, we apply the microwave drive at a frequency near the frequency of maximum response and characterize the increase in magnon density at the $f_-$ ESR frequency by measuring the NV relaxation rate $\Gamma_-$ between the 0 and –1 spin states (Fig. \ref{fig:figure3}b-c) (Supporting Information Note \ref{Sup:RateMeas}). Under the near-FMR driving, $f_-$ frequency magnons are added to the magnon gas as the scattering products of magnon-magnon interactions, resulting in an enhanced $\Gamma_-$. We measure $\Gamma_-$ at several drive frequencies (Fig. \ref{fig:figure3}d), as the location of maximum NV response changes slightly with drive power (Fig. \ref{fig:figure3}a). For all drive frequencies, we observe a strong increase in the relaxation rate for increasing drive power, reaching up to ~60 times its equilibrium value. Consistent with previous observations\cite{mccullian2020broadband} this process is strongly non-linear, as can be seen from the threshold power required to increase the relaxation rate at the higher drive frequencies.

If the magnon density is described by the Rayleigh-Jeans distribution, then we can determine the chemical potential by measuring the NV relaxation rates using\cite{du2017control}:
\begin{equation}
    \mu = h f_-\left(1-\frac{\Gamma_-(0)}{\Gamma_-(\mu)}\right)
\end{equation}
where $\Gamma_-(0)$ is the relaxation rate in the absence of microwave driving and $\Gamma_-(\mu)$ is the relaxation rate measured at a raised chemical potential caused by driving coherent spin waves. A key characteristic of the chemical potential for a bosonic system is that its maximum value is set by the bottom of the energy band\cite{du2017control, pitaevskii2016bose}, which in our system is located about 400 MHz below the FMR (at 20 mT) as can be calculated from the spin-wave dispersion (Supporting Information Note \ref{Sup:RateCalc}). Using Eq. 1 to calculate the chemical potential from the measured NV relaxation rates (Fig. \ref{fig:figure3}d), we find values far above the FMR, thereby exceeding this maximum (Fig. \ref{fig:figure3}e). We therefore conclude that the magnon gas created by near-FMR driving cannot be described by the Rayleigh-Jeans distribution with a finite chemical potential. Presumably, the magnon density is instead concentrated in a finite frequency range near the bottom of the spin-wave band that includes our detection (ESR) frequency. The strong increase in magnon density, compared to that observed in thinner YIG films\cite{du2017control}, might be related to the lower threshold power needed for triggering non-linear spin-wave responses in thicker magnetic films\cite{lee2020nanoscale}. Spectroscopic techniques such as Brillouin Light Scattering\cite{demokritov2006bose, kabos1994measurement} could shed further light on the spectral characteristics of the out-of-equilibrium magnon gas.

\begin{figure}[H]
    \centering
    \includegraphics[width=\textwidth]{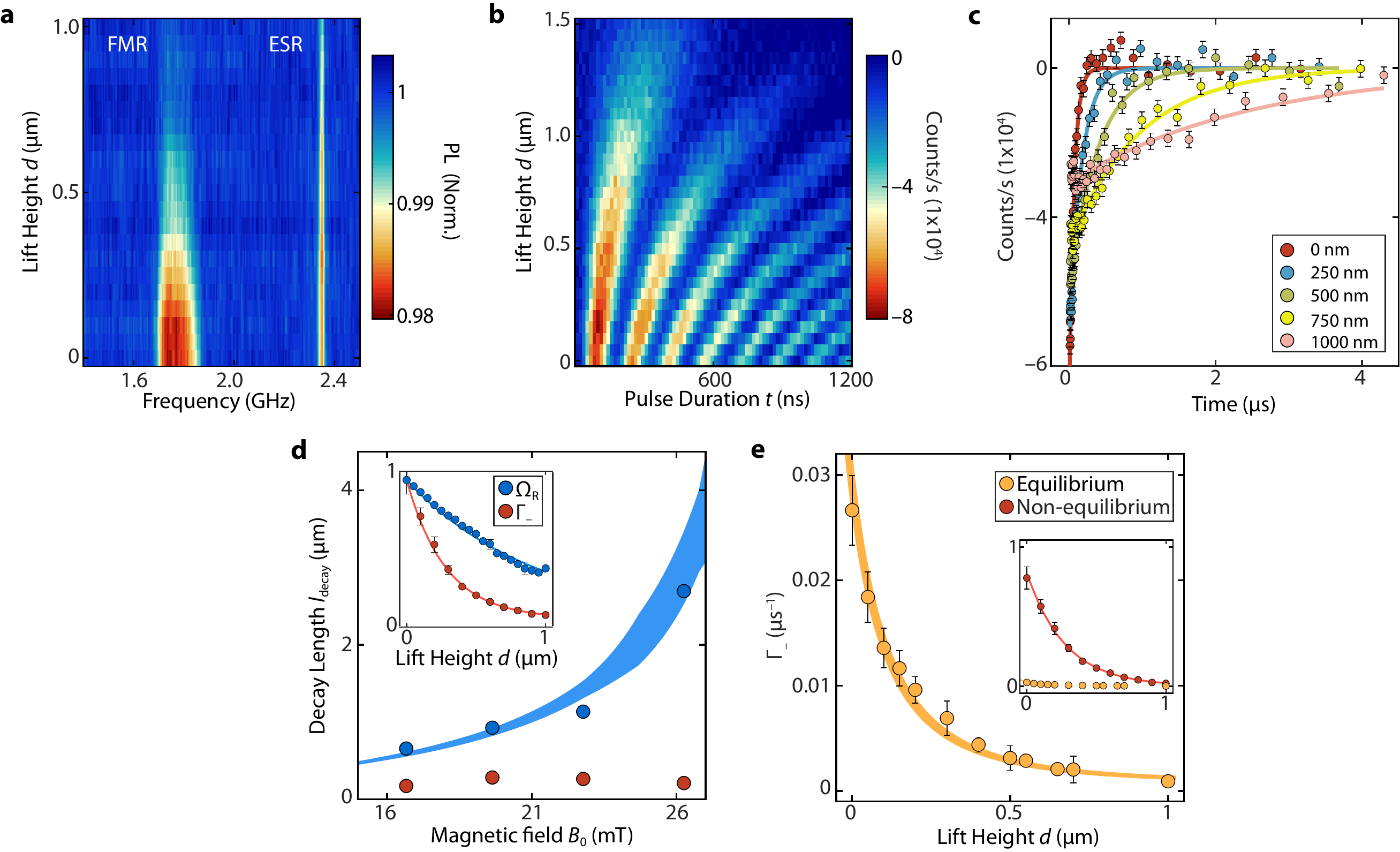}
    \caption{\textbf{Characterizing the wavenumber content of the equilibrium and out-of-equilibrium magnon gases.} (a) Normalized NV photoluminescence (PL) vs tip lift height $d$ and microwave drive frequency, measured at $\sim$\SI{150}{\micro\meter} to the right of the stripline. The decay of the signal for increasing lift height is caused by the decay of the spin-wave stray fields driving the NV spins. The tip touches the sample at \SI{0}{\micro\meter}. (b) Coherent NV spin rotations (Rabi oscillations) vs lift height $d$. The rotations are driven by a coherent spin-wave pulse of varying duration launched by the stripline at the $f_-$ ESR frequency at $B_0 =$ 20 mT. The measurement was performed at \SI{37}{\micro\meter} to the right of the stripline. (c) NV spin relaxation measurements for different tip lift heights. $B_0 = 20$ mT. Filled circles: data. Solid lines: fits to PL $= Ae^{(-2\Gamma_{-} t)}$. (d) Measured spatial decay length of the stray field of the driven coherent spin waves (blue circles) and of the out-of-equilibrium magnon gas (red circles) vs $B_0$. The blue band indicates the expected decay length for a coherent spin wave that is resonant with the NV ESR frequency, determined by calculating the spin-wavelength from the spin-wave dispersion while considering an uncertainty of $\pm 10$ nm in film thickness\cite{bertelli2020magnetic}, the $\pm$\SI{1.8}{\degree} uncertainty in the angle between the diamond cantilever and the YIG surface, and the uncertainty in the angle of $B_0$ (Supporting Information Note \ref{Sup:Bfield}). Inset: Spatial decay of the NV Rabi frequency ($\Omega_{\mathrm{Rabi}}$) when the Rabi oscillations are driven via coherent spin waves (blue circles) and of the NV relaxation rate ($\Gamma_-$) caused by the out-of-equilibrium magnon gas (red circles) at $B_0=20$ mT. Solid lines: fits to $Ae^{-d/l_{\mathrm{decay}}}$. (e) NV relaxation rate vs lift height $d$ without microwave drive (orange circles). Orange line: calculation for an equilibrium occupation of the spin-wave band. The width of the line considers the uncertainties discussed in (e). Inset: Comparison of the spatial decay of the NV relaxation rates without (as in the main panel, orange circles) and with microwave drive (as in inset in (d), red circles) near the FMR. $B_0=20$ mT. Red line: fit to $\Gamma_- = Ae^{-d/l_{\mathrm{decay}}}$. The two lines possess a similar decay constant.}
    \label{fig:figure4}
\end{figure}

We further characterize the out-of-equilibrium magnon density by probing its spatial-frequency content. To do so, we measure the spatial decay of the spin-wave stray fields away from the film, which is determined by the spatial frequencies (wavenumbers) of the magnons that generate the fields\cite{rustagi2020sensing}. We observe that the stray fields associated with the incoherent magnon gas decay much more rapidly with increasing NV-film distance than the stray fields generated by the coherently driven spin waves at the NV ESR frequency (Fig. \ref{fig:figure4}a). To quantify this difference, we first characterize the decay of the stray field $B_{\mathrm{SW}}$ of a coherent spin wave with a single, well-defined wavenumber $k_{\mathrm{SW}}$ that we excite by applying a microwave drive resonant with the $f_-$ NV ESR frequency using the stripline. The amplitude of this field decays exponentially with distance $d$ according to\cite{bertelli2020magnetic}: 
\begin{equation}
    B_{\mathrm{SW}} \propto e^{(-k_{\mathrm{SW}}d)}
\end{equation}
Because the excitation frequency is resonant with the NV ESR frequency, the field $B_{\mathrm{SW}}$ drives coherent NV spin rotations (Rabi oscillations) with a rotation rate (Rabi frequency, $\Omega_R$) that is proportional to the stray-field amplitude\cite{bertelli2020magnetic}: $\Omega_{\mathrm{R}} \propto B_{\mathrm{SW}}$. 

To quantify the decay length, we measure the NV Rabi frequency, $\Omega_{\mathrm{R}}$ as a function of the tip-sample distance (Fig. \ref{fig:figure4}b). By fitting the spatial decay using $\frac{\Omega_{\mathrm{R}}}{\Omega_{\mathrm{R}}(d=0)} =e^{(-k_{\mathrm{SW}} d)}$, we extract wavenumber, $k_{\mathrm{SW}}$ of the spin waves and the corresponding decay length, $l_{\mathrm{decay}}$ which ranges between $\sim0.65$ and \SI{2.7}{\micro\meter} depending on the external field $B_0$ (Fig. \ref{fig:figure4}d, blue dots). We find a good agreement with the wavenumber calculated from the spin-wave dispersion (Fig. \ref{fig:figure4}d, blue filled area, Supporting Information Note \ref{Sup:RateCalc}), demonstrating the power of height-dependent measurements for determining spatial frequencies.

We find that the stray fields of the out of-equilibrium magnon gas, generated upon driving near the FMR, decay on a much shorter length scale i.e., $\sim$ \SI{280}{\nano\meter} at $B_0 = 20$ mT (Fig. \ref{fig:figure4}a). To quantify the corresponding decay length, we measure the NV relaxation rate $\Gamma_-$ at different tip-sample distances $d$ (Fig. \ref{fig:figure4}c). By fitting the spatial decay of the relaxation rate using an exponential approximation $\frac{\Gamma_-}{\Gamma_-(d=0)} = e^{-d/l_{\mathrm{decay}}}$, we observe that the decay length $l_{\mathrm{decay}}$ is below \SI{1}{\micro\meter} over the entire range of $B_0$ (Fig. \ref{fig:figure4}d, red dots). This short decay length contrasts with that measured for the coherent spin waves (Fig. \ref{fig:figure4}d, blue dots), reflecting the additional presence of large-wavenumber magnons in the incoherent magnon gas. 

To examine if this is different for a magnon gas in equilibrium, we compare the NV relaxation rates measured in the absence of microwave driving to a calculation of the stray-field noise generated by a magnon gas in thermal equilibrium with zero chemical potential (Fig. \ref{fig:figure4}e). This calculation is based on a model\cite{rustagi2020sensing} that assumes a Rayleigh-Jeans occupation of the spin-wave band and calculates the stray-field noise at the NV ESR frequency by summing the contributions of all spin-wave modes at this frequency (Supporting Information Note \ref{Sup:RateCalc} \& Fig. \ref{fig:figure4}e, orange line). This model was recently demonstrated to accurately describe the stray-field noise of thin magnetic films\cite{rustagi2020sensing}. We find a quantitative match with the measured equilibrium NV relaxation rate $\Gamma_-(\mu=0)$ if we assume a $0.28 \pm$ \SI{0.3}{\micro\meter} distance offset of the NV centers at zero tip-lift height (Fig. \ref{fig:figure4}e). This offset is larger than the NV implantation depth of $\sim20$ nm, which could be caused by small particles picked up by the tip during scanning. Last, we compare the measured relaxation rate under near-FMR driving to the same model but scaled by a prefactor. We find a similar exponential decay of the NV relaxation rate with increasing distance to the film as for the equilibrium case (inset Fig. \ref{fig:figure4}e), indicating a similar spatial-frequency content of the equilibrium and out-of-equilibrium magnon gases. Furthermore, the calculations confirm that the spatial decay length should not depend strongly on the external field $B_0$ (Supporting Information Note \ref{Sup:RateCalc}), consistent with the measurements shown in Fig. \ref{fig:figure4}d.

We have shown that coherent spin waves enable the generation of a high-density magnon gas unidirectionally with respect to an excitation stripline. The threshold power required to trigger this process underscores the non-linearity of the underlying magnon scattering. From the more than 10-fold increase of the stray-field noise under near-FMR driving, probed via relaxometry measurements of our sensor spin, we conclude that the resulting magnon gas cannot be described by a Rayleigh-Jeans occupation of the spin-wave band. We demonstrate that the spatial decay length of the spin-wave stray fields contains valuable information about the spatial frequencies of the spin waves generating the fields. The observed sub-micron spatial decay lengths of the stray fields generated by the out-of-equilibrium magnon gas indicate the presence of small-wavenumber magnons and highlights the need for proximal sensors such as the scanning-probe NV magnetometer. Further controlling the directionality of the excited coherent spin waves by e.g. shaping stripline geometries and/or tuning the direction of the magnetic external field could enable delivering high-density magnon gases to target locations in a magnetic film or device. Targeted delivery of high-density magnons gases provides new opportunities for controlling spin transport and for triggering magnetic phenomena such as phase transitions\cite{nan2020electric}, magnon condensation\cite{schneider2020bose, demokritov2006bose,demokritov2008quantum}, and spin-wave-induced domain-wall motion\cite{kim2014propulsion, shen2020driving, chang2018ferromagnetic,wang2015magnon}.

\subsection*{Supporting Information}
Experimental and theoretical details on the YIG sample, the measurement setup, the fabrication of the scanning tip, the calibration of the external magnetic field magnitude and direction, the theoretical computation of the NV relaxation rate induced by thermal magnons and the NV relaxation rate measurement are included in the Supporting Information. 

\section*{Acknowledgements}\addcontentsline{toc}{subsection}{Acknowledgements}
\textbf{Funding:} This work was supported by the Dutch Research Council (NWO) through the NWO Projectruimte grant 680.91.115 and the Kavli Institute of Nanoscience Delft.\hfill\break
\textbf{Author contributions:} B.G.S, S.K., A.K. and T.v.d.S. conceived and designed the experiments and realized the imaging setup. B.G.S., S.K., and A.K performed the experiments. B.G.S., S.K., T.v.d.S. and H.L. analyzed and modelled the experimental results with contributions of I.B. and J.J.C. I.B. fabricated the stripline on the YIG sample. B.G.S, M.R, N.d.J. and H.v.d.B. fabricated the diamond cantilevers. B.G.S., S.K., and T.v.d.S wrote the manuscript with contributions from all coauthors. \hfill\break
\textbf{Competing interests:} The authors declare that they have no competing interests. \hfill\break
\textbf{Data and materials availability:} All data contained in the figures will be made available at Zenodo.org upon publication. Additional data related to this paper may be requested from the authors. 

\phantomsection\addcontentsline{toc}{subsection}{References}
\printbibliography
\end{refsection}
\newpage
\beginsupplement 
\begin{refsection} 
\textbf{{\huge Supporting Information}}\hfill\break
\phantomsection
\addcontentsline{toc}{section}{Supporting Information}
\localtableofcontents
\subsection{YIG Sample}\label{Sup:YIG}
We study a ~235 ± 10 nm thick film of (111)-oriented ferrimagnetic insulator yttrium iron garnet (YIG) grown on a gadolinium gallium garnet substrate by liquid-phase epitaxy (Matesy GmbH). The stripline (length of 2 mm, width of \SI{30}{\micro\meter}, and thickness of 200 nm) used for spin-wave excitation was fabricated directly on the YIG surface using e-beam lithography, using a double layer PMMA resist (A8 495K / A3 950K) and a top layer of Elektra95, followed by the deposition of 5 nm / 200 nm of Cr/Au.

\subsection{Measurement setup}\label{Sup:MeasSetup}
Our scanning NV-magnetometry setup is equipped with two stacks of Attocube positioners (ANPx51/RES/LT) and scanners (ANSxy50/LT and ANSz50/LT) that allow for individual positioning of the tip and sample. We first position the NV tip in the focus point of the objective (LT-APO/VISIR/0.82) and use the scanners of the sample stack to create spatial PL images (lateral and height). To record the NV PL, we use our home-built confocal setup, which is equipped with a 515 nm green laser (Cobolt 06-MLD, pigtailed) for NV excitation. We use a fiber collimator (Schäfter+Kirchhoff 60FC-T) to couple the laser out into free space. A dichroic mirror (Semrock Di03-R532-t3-25x36) is used to reflect the green laser, which is then focused by the objective lens (LT-APO/VISIR/0.82) on the tip of a homemade all-diamond cantilever probe (Supporting Information Note \ref{Sup:TipFab}). The shape of the tip aides the guiding of the NV PL back towards the objective where it is collimated and transmitted by the dichroic mirror and additionally filtered by a long-pass filter (BLP01-594R-25) before it is focused on the chip of an avalanche photodiode (APD) (Excelitas SPCM-AQRH-13). The resulting signal is collected and counted by a National Instruments DAQ card. A SynthHD (v2) dual channel microwave generator (Windfreak Technologies, LLC) is used for driving the NVs and exciting the spin waves. High-speed pulse sequences are generated by a PulseBlasterESR-PRO pulse generator (SpinCore Technologies, Inc.).

\subsection{Diamond tip fabrication}\label{Sup:TipFab}
\subsubsection{NV implantation}
Our scanning tip is fabricated from a (001)-oriented electronic grade type IIa diamond grown via chemical vapor deposition by Element 6. The diamond is laser cut and polished (Almax EasyLab) into 2x2x0.05 mm$^3$ chips and subsequently cleaned in fuming nitric acid. To remove surface damage from the polishing \cite{appel2016fabrication}, we use inductively coupled plasma reactive ion etching (ICP-RIE) to remove the top $\SI{5}{\micro\meter}$. The diamond is implanted with 15N ions at \SI{6}{\kilo\electronvolt} with a dose of 10$^{13}$ ions/cm$^2$ by INNOViON. After implantation, we clean the samples in a tri-acid solution (H$_2$SO$_4$:HClO$_4$: HNO$_3$ = 1:1:1) at \SI{120}{\degreeCelsius} for 1 hour. We then anneal the diamond for 8 hours at \SI{800}{\degreeCelsius} at approximately 10$^{-6}$ mbar, followed by another cleaning step in a tri-acid solution.

\subsubsection{Structuring diamond} 
To structure the diamond into a scanning tip and for mounting this tip in our AFM setup, we follow Refs. \cite{maletinsky2012robust,appel2016fabrication}. We first thin down the $\sim$\SI{45}{\micro\meter} thick diamond to $\sim$5-\SI{7}{\micro\meter} in a 1.2x1.2 mm$^2$ area in the center \cite{appel2016fabrication, ruf2019optically}. The diamond is then flipped with the NV side facing up and glued (with PMMA) to a silicon carrier wafer to enable spin coating. An 40-50 nm titanium etch mask is deposited using RF-magnetron sputtering and a layer of about 800 nm PMMA (950K A8) is spin coated and patterned using e-beam lithography into 20x\SI{50}{\micro\meter}$^2$ rectangular masks. The mask is then transferred into the titanium using SF6/He plasma RIE. Subsequently, an anisotropic oxygen etch transfers the cantilevers into the diamond with a thickness of about \SI{3}{\micro\meter} \cite{appel2016fabrication, ruf2019optically}. We clean the sample using hexafluoride (40\%, HF) and deposit a new layer of 5 nm of titanium to improve the adhesion of the FOx-16 resist that we spincoat subsequently and which serves as the etch mask for the \SI{1.5}{\micro\meter} tall diamond pillars that are created in a final anisotropic oxygen etch. The resulting tips have a diameter of approximately 300 nm and therefore contain about 100 NVs at 20 nm below its apex.
The diamond is cleaned to get rid of the titanium/FOx mask in HF and in fuming nitric acid (100\%) to remove possible organic contaminants. 
Finally, the diamond tip is glued (UV glue) to the end of a pulled optical fibre that is connected to a tuning fork for AFM operation\cite{appel2016fabrication}.  

\subsection{Magnetic field calibration}\label{Sup:Bfield}
In our experiments we apply the magnetic external field $\vec{B_0}$ using a permanent magnet (S-10-20-N, Supermagnete) onto a linear translation stage (MTS25-Z8, Thorlabs, controlled by a KDC101 Thorlabs servo motor). To vary the strength of the external field, we change the position of the stage to bring the magnet closer to the sample and tip. We orient the magnet in such a way that the field aligns with one of the four NV families\footnote{with 'NV family', we mean the collection of NV centers with the same crystallographic orientation} (which is $54.7 \pm 1.8^\circ$ with respect to the normal of the YIG surface) and to magnetize the YIG film along the length of the stripline. At each magnetic field, we calibrate its magnitude and orientation by measuring the eight ESR frequencies $f_{i = 1-8}$ of the four NV families($\mathrm{NV}_{1-4}$) (Fig. \ref{fig:figure1}b, main text). From these frequencies we determine $\vec{B}_0$ by performing a least square minimization\cite{lillie2020Bfieldfit}:
\begin{equation}
    \min(\sum_{i=1}^8(f_i-f_i^\mathrm{calc}(\vec{B}))^2
    \label{least_sqaures}
\end{equation}
where $f_i^\mathrm{calc}$ are the eight ESR frequencies calculated from the NV Hamiltonian at a magnetic field $\vec{B}$ \cite{van2015nanometre}:
\begin{equation}
    H = DS_\mathrm{Z}^2 + \gamma\vec{B}\cdot\vec{S}.
    \label{NV_H}
\end{equation}
\begin{figure}[H]
    \centering
    \includegraphics[width=\textwidth]{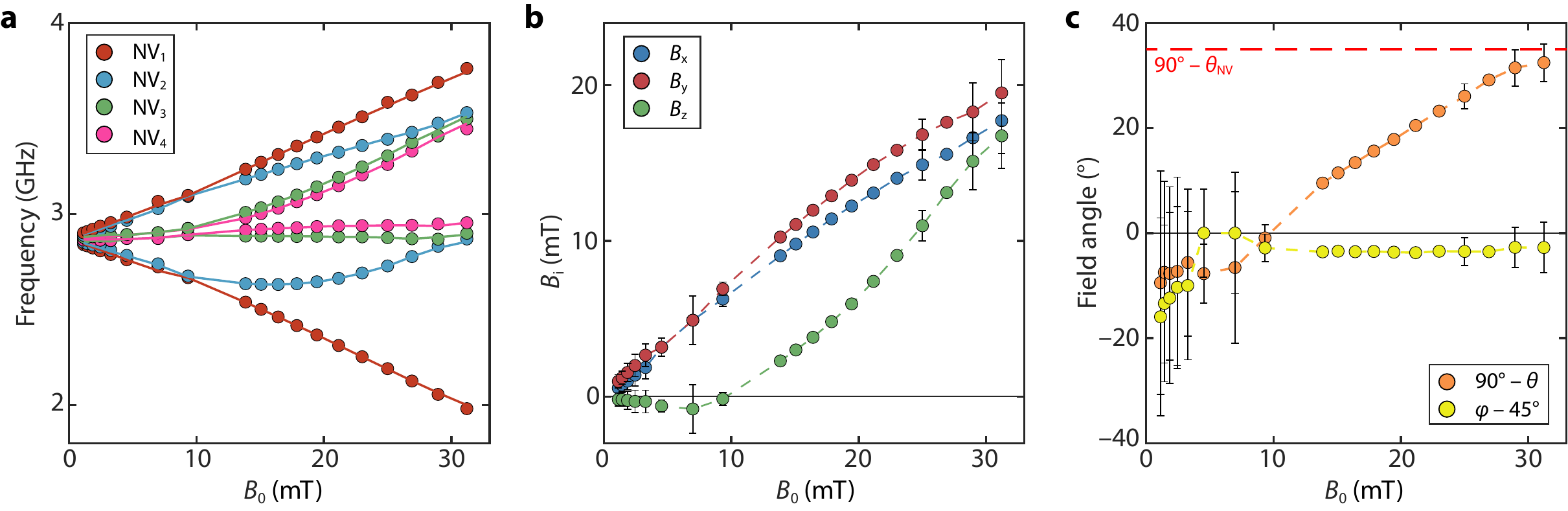}
    \caption{Vector magnetic field determination. (a) The eight ESR frequencies as a function of applied external magnetic field ($B_\text{0}$). Filled circles: data. Solid lines: fitted ESR frequencies obtained by performing a least square minimization of Eq. \eqref{least_sqaures} at each field. (b) The magnetic field components ($B_\text{x}$, $B_\text{y}$, $B_\text{z}$) as a function of the field magnitude $B_0$. The error bars represent 95\% confidence bounds. (c) The magnetic field angles as a function of $B_\text{0}$. The error bars represent 95\% confidence bounds.}
    \label{fig:magnet_calibration}
\end{figure}
Here, $D$ is the zero-field splitting (2.87 GHz), $S_\mathrm{i = X,Y,Z}$ are the Pauli spin-1 matrices, and $\gamma$ is the electron gyromagnetic ratio (28 GHz/T). Capital XYZ denote the NV frame. The Z-axis is taken along the NV axis and thus differs for the four NV families, pointing along the unit vectors
\begin{equation}
    \hat{n}_\mathrm{NV}=\frac{1}{\sqrt3}({\pm}1,{\pm}1,1)
    \label{unit_vector}
\end{equation}
expressed in the lab/diamond frame. The microwave stripline is oriented along the [110] direction in this frame.

After fitting the eight ESR frequencies, $f_{i=1-8}$, (Fig. \ref{fig:magnet_calibration}a),we extract the magnetic field components ($B_\mathrm{x}$, $B_\mathrm{y}$, $B_\mathrm{z}$) (Fig. \ref{fig:magnet_calibration}b), the magnetic-field magnitude $B_\mathrm{0}$ and the corresponding out-of-plane angle, $\theta$, and in-plane angle, $\varphi$ (Fig. \ref{fig:magnet_calibration}c). Lowercase xyz refer to the lab frame.

The field is oriented in-plane up to 10 mT (see $B_\text{z}$, Fig S.1b), after which the out-of-plane component is increased such that the field matches the target NV axis at 30 mT (red dashed line in Fig. \ref{fig:magnet_calibration}c). This ensures a good $f_-$ ESR contrast over the entire field range \cite{tetienne2012NV}.

\subsection{NV relaxation induced by thermal magnons}\label{Sup:RateCalc}
We follow the approach of Rustagi et al.\cite{rustagi2020sensing} to calculate the NV relaxation rates induced by the magnons in our YIG film, using
\begin{align}
    \Gamma_\mp (\omega_\mp) 
        = 
        \frac{\gamma^2}{2} \int \frac{d\vec{k}}{(2\pi)^2} \sum_{{i,j} \in \{x,y\}} \mathcal{D}^\text{eff}_{\pm i}(\vec{k}) \mathcal{D}^\text{eff}_{\mp j}(-\vec{k}) C_{ij} (\vec{k}, \omega_\mp).
    \label{eq:NV_relaxation_rate}
\end{align}
Here, $\Gamma_{\mp}$ are the relaxation rates corresponding to the $\omega_{\mp}$ ESR  frequencies, $\vec{k}$ is the spin-wavevector,  $\mathcal{C}$ is a spin-spin correlator describing the thermal magnon fluctuations, and $\mathcal{D}^\text{eff}$ is a dipolar tensor that calculates the magnetic stray fields that induce NV spin relaxation generated by these fluctuations. We will now summarize how these quantities are calculated for our measurement geometry. In addition, we will discuss the expected distance dependence of the relaxation rate that we compared with experiments in Fig. 4e of the main text. 

The thermal transverse spin fluctuations in the film are described by \cite{rustagi2020sensing}: 
\begin{align}
    C_{ij} (\vec{k}, \omega) = 
        2D_{th}\sum_{\nu = \{x,y\}}S_{i\nu}(\vec{k}, \omega)S_{j\nu}(-\vec{k}, -\omega)
            \label{eq:correlations}
\end{align}
where $D_\text{th} = \frac{\alpha k_B T}{\gamma M_\text{s} L}$, with $k_B$ the Boltzmann constant, $T$ the temperature, and
\begin{align}
    S (\vec{k}, \omega)
        =
        \frac{\gamma}{\Lambda}
        \begin{bmatrix}
            \omega_3 - i\alpha \omega & -\omega_1 - i\omega \\
            -\omega_1 + i\omega       & \omega_2 - i\alpha \omega 
        \end{bmatrix}
    \label{eq:susceptibility_matrix}
\end{align}
is the spin-wave susceptibility, with\cite{rustagi2020sensing}  
\begin{align}\label{eq:omegas}
    \omega_0 (\vec{k})
        &=
        \omega_\text{B} \cos(\theta_0 - \theta) - \omega_\text{M} \cos^2 \theta_0 + \omega_D k^2, \\
    \omega_1 (\vec{k})
        &=
        \omega_\text{M} f_k \sin \phi_k \cos \phi_k \cos \theta_0, \\
    \omega_2 (\vec{k})
        &=
        \omega_0 + \omega_\text{M} \left[ f_k \cos^2 \phi_k \cos^2 \theta_0 + (1- f_k) \sin^2 \theta_0 \right], \\
    \omega_3 (\vec{k})
        &=
        \omega_0 + \omega_\text{M} f_k \sin^2 \phi_k, \\
    \Lambda (\omega)
        &=
        (\omega_2 - i\alpha \omega)(\omega_3 - i\alpha \omega) - \omega_1^2 - \omega^2.
\end{align}
Here, $f_k \equiv 1-(1-e^{-kL})/(kL)$ with $L$ the film thickness, and $\phi_k$ is the polar angle of a spin wave in $k$-space. The spin-wave dispersion is obtained by taking the real part of the solutions of $\Lambda =0$. To calculate $S$ we use $L =  235 \pm 10$ nm, Gilbert damping $\alpha = (1.2 \pm 0.1) \cdot 10^{-4}$, $M_\mathrm{s} = (1.42 \pm 0.01) \cdot 10^5$ A/m \cite{bertelli2020magnetic} and  $A_\text{ex} = (3.7 \pm 0.4) \cdot 10^{-12}$ J/m \cite{klingler2014measurements}. The equilibrium angle of the magnetization $\theta_0$ is obtained by finding the minimum of the free energy at each value of the magnetic field ($\theta_0$ is in-plane to within a few degrees for the field range used in our measurements).

The dipolar tensor $D^\text{eff}(\vec{k},\omega)$ calculates the magnetic stray fields that induce NV spin relaxation generated by the thermal magnons in the film. Because the correlator $C$ is expressed in the frame of the magnet (with a z-axis pointing along the equilibrium magnetization), $D^\text{eff}(\vec{k},\omega)$ is obtained by first rotating the magnet frame to the lab frame, then multiplying by the dipolar tensor $\mathcal{D}(\vec{k})$ in the lab frame, and then rotating the result to the NV frame: $D^\text{eff}(\vec{k},\omega) = R_{yz}(\theta_\text{NV}, \phi_\text{NV}) \mathcal{D}(\vec{k})R_Y(\theta_0)^T$, where 
\begin{align}
    \mathcal{D}(\vec{k})
        =
        -\frac{\mu_0 M_s}{2} e^{-|\vec{k}| d_\text{NV}} (1 - e^{-|\vec{k}|L})
        \begin{bmatrix}
            \cos^2 \phi_k & \sin(2\phi_k)/2 & i\cos \phi_k \\ 
            \sin(2\phi_k)/2 & \sin^2 \phi_k & i\sin \phi_k \\
            i\cos \phi_k & i\sin \phi_k & -1
        \end{bmatrix},
    \label{eq:dipolar_tensor_in_lab_frame}
\end{align}
where $\mu_0$ is the vacuum permeability and $d_\text{NV}$ is the distance between the NV and the sample surface. The terms in Eq. \eqref{eq:NV_relaxation_rate} that induce spin relaxation are given by\cite{rustagi2020sensing}: $\mathcal{D}^\text{eff}_{\pm \nu} = \mathcal{D}^\text{eff}_{x \nu} \pm i\mathcal{D}^\text{eff}_{y \nu}$.

\begin{figure}[ht]
    \centering
    \includegraphics[width=.4\textwidth]{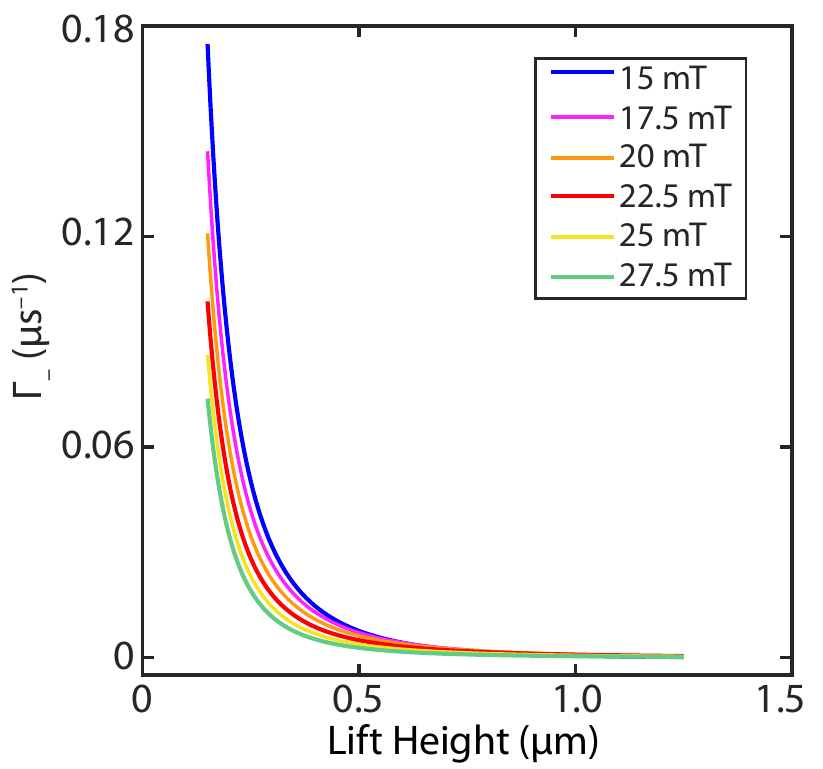}
    \caption{Calculated relaxation rate as a function of tip lift height for an in-equilibrium occupation of the spin-wave band for different magnetic fields. We used $\theta_\text{NV} = 54.7^\circ$ and field-angle $\theta=71.7^\circ$ with respect to the surface normal.}
    \label{fig:rate_distance_fields}
\end{figure}

For a magnon gas in thermal equilibrium, in the absence of microwave driving, the dependence of the NV relaxation rate on the NV-sample distance can be calculated using Eq. \eqref{eq:NV_relaxation_rate}. We find a good match between the calculated and measured rate (Fig. 4e main text) if we include an offset distance of $0.28 \pm 0.03$ \si{\micro\meter}. This offset distance is larger than the ~20 nm NV implantation depth, which could be caused by small particles picked up by the tip during scanning. We observe a similarly fast decay over a broad range of magnetic field values (Fig. \ref{fig:rate_distance_fields}).  

\subsection{Extracting the NV relaxation rate}\label{Sup:RateMeas}
In this section, we describe how we obtain the relaxation rate of a target NV family, plotted in Figs. \ref{fig:figure3} and \ref{fig:figure4} of the main text\footnotemark[1]. To extract the $\Gamma_-$ relaxation rate of the target NV family we apply the pulse sequence shown in Fig. \ref{fig:figure3}b of the main text. We perform two sequences: without and with a microwave $\pi$-pulse on the $\ket{0} \leftrightarrow \ket{-1}$⟩ ESR transition of the target NV family. For the measurement without a $\pi$-pulse, we write the photoluminescence collected during readout as:
\begin{equation}
N(t) = p_0(t)N_0 + p_{-1}(t)N_{-1} + p_{+1}(t)N_{+1} + N_\mathrm{BG}  
\label{PL_total}
\end{equation}
where $N_{i = -1, 0, +1}$ are the number of collected photons when the target NVs are in state ${i}$ and $p_{i = -1, 0, +1}$ are the associated occupation probabilities. $N_\mathrm{BG}$ is the background PL, which includes the contribution of the other NV families.

Applying a $\pi$-pulse on the $\ket{0} \leftrightarrow \ket{-1}$⟩ transition of the target NV family switches the populations of these states. The photoluminescence collected during readout now is
\begin{equation}
N_\pi(t) = p_{-1}(t)N_{0} + p_{0}(t)N_{-1}+ p_{+1}(t)N_{+1}+ N_\mathrm{BG}.  
\label{PL_pi}
\end{equation}
By taking the difference of equations \eqref{PL_total} and \eqref{PL_pi}, the background contribution drops out, giving
\begin{equation}
N - N_\pi = (p_0(t) - p_1(t))(N_0 - N_{-1}).
\label{meas_difference}
\end{equation}
For the field range used in our experiments $\Gamma_+\ll\Gamma_-$ \cite{rustagi2020sensing} because the $\omega_+$ transition is far detuned from the FMR and moreover less affected by the fields produced by the spin waves in the YIG film. As such, the time dependence of $N - N_\pi$ is dominated by the $\Gamma_-$ relaxation rate and follows an exponential decay $N - N_\pi = A e^{-2\Gamma_- t}$ that we use to extract $\Gamma_-$. 
\printbibliography
\end{refsection}
\end{document}